%% file: main_12_arxiv.tex
\documentclass[final,1p,authoryear]{elsarticle}
\usepackage[colorlinks=true,linkcolor=blue,citecolor=blue,urlcolor=blue]{hyperref}
\usepackage{doi}
\usepackage{url}

\usepackage[english]{babel}
\usepackage{csquotes}
\usepackage{amsmath,amssymb,amsfonts}
\usepackage{graphicx} 
\usepackage[dvipsnames]{xcolor} 
\usepackage{siunitx} 
\usepackage[toc,page]{appendix} 
\usepackage[normalem]{ulem}
\usepackage{bm} 
\usepackage{soul} 
\usepackage{cancel} 
\usepackage{braket}

\newcommand{\SubFig}[2]{\ref{#1}{\color{blue}#2}}

\usepackage{times}
\usepackage{latexsym}
\usepackage[running,mathlines]{lineno} 
\usepackage{verbatim} 
\usepackage{bm} 
\usepackage{footnote} 

\usepackage{setspace} 
\makeatletter
\renewcommand\normalsize{\@setfontsize\normalsize{11pt}{16pt}} 
\makeatother 

\usepackage[round,authoryear]{natbib} 
\newcommand{\parencite}{\citep}
\newcommand{\textcite}{\citet}  

\journal{European Journal of Operational Research} 

\begin{document}

\begin{frontmatter}

\title{Optimization Algorithm for Inventory Allocation in Gravity-Flow Racks with Classical and Quantum-Hybrid Computing}

\author[inst1]{Gabriel P. L. M. Fernandes\corref{cor1}}
\ead{gabrielpedro@df.ufscar.br}
\cortext[cor1]{Contact author}

\author[inst1]{Matheus S. Fonseca}
\ead{msfonseca@estudante.ufscar.br}

\author[inst1]{Amanda G. Valério}
\ead{amanda.valerio@estudante.ufscar.br}

\author[inst1]{Alexandre C. Ricardo}
\ead{alexandre.ricardo@df.ufscar.br}

\author[inst1]{Nicolás A. C. Carpio}
\ead{ncarpio@estudante.ufscar.br}

\address[inst1]{Departamento de Física, Universidade Federal de São Carlos, 13565-905, São Carlos, São Paulo, Brazil}

\author[inst2]{Paulo C. C. Bezerra}
\ead{bezerra@wvblabs.com.br}
\address[inst2]{Wernher von Braun Advanced Research Center, 13098-392, Campinas, São Paulo, Brazil}

\author[inst1]{Celso J. Villas-Boas}
\ead{celsovb@df.ufscar.br}

\begin{abstract}
Warehouses play a central role in industrial logistics, functioning as critical hubs for storing and organizing inventory to support efficient production. Optimizing item allocation within these facilities is essential for reducing operational costs and improving delivery times. In this work, we address the optimization of inventory allocation in warehouses equipped with gravity-flow racks, which are designed for First In, First Out (FIFO) logistics, a configuration that inherently requires item reinsertions during retrieval operations to maintain flow continuity. These reinsertions, however, are time-consuming and costly, so minimizing their occurrence is crucial for operational efficiency. We propose an optimization strategy that simultaneously allocates multiple items, determining their placement across available shelves in a single decision step, explicitly accounting for every item and every shelf in the warehouse. By jointly evaluating multiple items, our approach enables globally optimized placement decisions, minimizing conflicts that arise in sequential methods. The problem is formulated as a Quadratic Unconstrained Binary Optimization (QUBO), allowing implementation on both classical metaheuristics and quantum-hybrid solvers. We assess performance by comparing three classical optimization approaches — two variants of Simulated Annealing and the commercial solver Gurobi — with D-Wave’s hybrid solver, which uniquely combines quantum annealing with classical metaheuristics.  Benchmark experiments demonstrate that, as problem size increases, the hybrid quantum-classical solver consistently produces high-quality solutions more efficiently than the classical metaheuristics. Complementing these benchmarks, a factory-scale simulation based on real operational data shows that considering larger batches of items in the allocation step can significantly reduce reinsertions, highlighting the practical potential of the proposed approach for industrial logistics.

\end{abstract}


\end{frontmatter}

\section{Introduction}

Effectively managing an industrial warehouse presents several challenges, among which inventory allocation stands out as a significant concern due to the continuous inflow and outflow of items required to meet operational demands~\parencite{Faber2013}. Consequently, the development of optimized allocation strategies remains a core objective in logistics and operations research, as efficient allocation can significantly reduce delivery times, improve forklift utilization, and lower overall operational costs~\parencite{tenHompel2007}.

To formulate an efficient strategy for optimizing item allocation in a warehouse, particularly focused on reducing operational costs associated with item-picking, one must consider the type of racking system employed, since each racking system entails its own operational conditions and practical requirements that influence how items can be effectively allocated. While numerous strategies have been proposed for warehouses equipped with multilevel racks~\parencite{Ene2011, Pan2014, Zhang2019}, relatively little attention has been paid to warehouses equipped with gravity-flow racks, despite their widespread use and proven advantages~\parencite{Richards2011}.

\begin{figure}
    \centering    \includegraphics[width=0.75\linewidth]{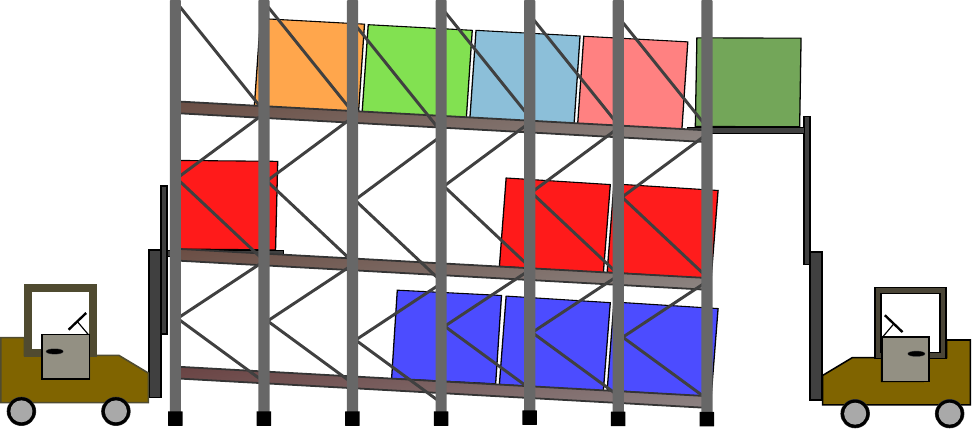} 
    \caption{Schematic representation of a gravity-flow rack. The rack has inclined shelves with rolling rails, allowing items to move forward under gravity. Items are represented by colored pallet boxes, and the rack comprises three gravity-flow shelves. On the left, at the loading side, a forklift places a red box onto the middle shelf. On the right, at the retrieval side, another forklift removes a green box from the top shelf. To reach an item located in the middle of a shelf (e.g., the dark green box on the top shelf), the forklift on the right must first remove all pallet boxes in front of it (e.g., the red and blue boxes), illustrating the operational constraints imposed by the FIFO method.}
    \label{fig:GravRacks}
\end{figure}

Gravity-flow racks, a type of racking system particularly appropriate for factory warehouses and distribution centers, are designed using slightly inclined shelves and rolling rails, allowing heavy items to slide forward under gravity ~\parencite{
Tompkins1998}. These systems are engineered to operate according to the \textit{First In, First Out} (FIFO) logistics method~\parencite{tenHompel2007}, which ensures that the items allocated to a shelf are retrieved in the order they were stored, leading to proper inventory rotation and reducing unnecessary storage time. A simplified depiction of a gravity-flow rack and the FIFO retrieval process are illustrated in Fig.~\ref{fig:GravRacks}. Compared to static, multilevel racks, gravity-flow systems provide a higher storage density by reducing the number of aisles needed for forklift transit~\parencite{Richards2011}, thereby maximizing storage space utilization and reducing forklift travel time.

Although these racking systems offer clear advantages, they also impose an often overlooked constraint stemming from their reliance on the FIFO method. When production lines request an item located on a specific shelf, forklifts must remove all items placed in front of it to reach the desired item~\parencite{tenHompel2007}. Items removed in this process that are not immediately needed must be reinserted into a shelf, not necessarily in their original location. While some reinsertions are expected, many of these operations can substantially increase operational costs, as each removal and reallocation requires additional forklift use and labor. 

This issue becomes even more problematic when ineffective allocation strategies repeatedly place items in far-from-optimal positions, triggering a cycle of avoidable reinsertions and unnecessary movement. In Fig.~\ref{fig:GravRacks}, for example, the items are arranged to illustrate an attempt at using dedicated shelving, a method where each shelf holds only a single item type, with each color representing a distinct item. However, if the number of distinct items exceeds the number of available shelves, this strategy fails, leading to a disorganized allocation in which unrelated items are forced to share the same shelf. Over time, such problems not only drive up costs but also disrupt operational flow, reducing overall warehouse efficiency and increasing the risk of workplace accidents.

To address these challenges, we propose an optimization strategy that, in a single decision step, determines the placement of multiple incoming items in gravity-flow rack warehouses to minimize the number of reinsertions required during retrieval operations. By assigning positions to several items simultaneously — rather than sequentially — the strategy enables better allocation decisions, reducing the risk of assigning an item to a location that would have been better suited to another item considered later. This approach naturally falls within the scope of combinatorial optimization, as it requires determining the optimal distribution of a set of $N$ items across $M$ gravity-flow shelves while satisfying (i) the capacity constraints of each shelf, (ii) the requirements imposed by the FIFO method, and (iii) the operational constraints of a real-world warehouse, where previously allocated items may already occupy part of the storage space. The resulting optimization task is computationally demanding, as the solution space grows combinatorially with respect to both items and shelves. This rapid growth places the problem within the class of NP-hard combinatorial optimization problems, for which, at even moderate scales, exact methods become infeasible and even state-of-the-art metaheuristics struggle to deliver consistent high-quality solutions.

Considering the advances that Quantum Computing has brought in other fields of mathematics by exploiting quantum phenomena to speed up specific tasks~\parencite{Montanaro2016}, quantum algorithms developed to address combinatorial optimization problems have been attracting increasing interest~\parencite{Yarkoni2022, Abbas2024}. In particular, Adiabatic Quantum Computing (AQC) has emerged as a promising quantum computing model for designing algorithms that tackle such problems~\parencite{Farhi2000, vanDam2001}. To search for high-quality solutions within the solution space of a given combinatorial optimization problem, AQC relies on gradually evolving a quantum system from the ground state (i.e., the lowest-energy state) of an Initial Hamiltonian, $H_I$, to the ground state of a Problem Hamiltonian, $H_P$, whose eigenstates encode the possible solutions of the optimization task. In this encoding, lower-energy eigenstates correspond to higher-quality solutions, with the ground state of $H_P$ representing the optimal one. Under sufficiently slow evolution, the probability of reaching the ground state of $H_P$ is close to 1~\parencite{Albash2018}, meaning that the optimal solution can be obtained with high probability.
 
Efforts to implement the described quantum computing model in real hardware led to the development of Quantum Annealing (QA), with successful implementations in D-Wave's annealing-based quantum computers~\parencite{Johnson2011}. These systems solve problems formulated as Quadratic Unconstrained Binary Optimization (QUBO) instances~\parencite{Glover2022, Boros2007}, and recent hardware improvements have enabled the implementation of proofs of concept in a growing number of industrial scenarios~\parencite{Yarkoni2022}. However, real-world scale problems still require a much larger number of quantum bits than currently available hardware can provide. To address this limitation, D-Wave introduced a class of hybrid solvers~\parencite{DWAVE2021, DWAVE2022}, which combine classical metaheuristics with Quantum Annealing to handle larger and more complex problems.

In this work, we formulate the strategy for optimizing item allocation in warehouses equipped with gravity-flow racks as a QUBO problem, enabling its implementation on classical, quantum, and quantum-hybrid hardware. The algorithms studied here were all developed by our team, and their solutions were compared with those obtained using a commercial software package Gurobi Optimizer \parencite{Gurobi}. We considered dimensions typical of an industrial plant, and for allocations involving a few hundred items, the solution provided by D-Wave’s hybrid system proved to be more efficient and accurate. The remainder of this work is organized as follows. In Sec.~\ref{SecMethods}, we provide an overview of Quantum Annealing, hybrid solvers, and Simulated Annealing (SA) – a classical metaheuristic that has influenced the development of quantum optimization methods and remains an effective benchmark for comparative analyses. In Sec.~\ref{SecModel}, we introduce the item allocation problem on gravity-flow racks, present our optimization strategy for minimizing reinsertions during retrieval operations, and define a Problem Hamiltonian that encodes this strategy. In Sec.~\ref{SecResults}, we report results from applying our strategy to distribute tens to hundreds of items across warehouse configurations with multiple shelves, comparing the performance of D-Wave’s hybrid solvers with two classical SA implementations — based on both binary and integer variable formulations — and with the commercial optimizer Gurobi. We analyze solution quality and computational runtime, and further report results from a simulation study based on real operational data from a factory comprising approximately 450 shelves of varying sizes, involving 170 distinct types of items and more than 50,000 insertion and removal operations over an approximately three-month period. Applying our strategy led to a reduction of more than 90\% in reinsertions relative to the system in use at the factory. In Sec.~\ref{SecConclusions}, we present our conclusions and outline directions for future research.

\section{Metaheuristics} \label{SecMethods}

Solving an optimization problem requires determining the combination of variables that minimizes an objective function while satisfying a set of constraints. Although such problems are particularly well-suited for modeling strategic decisions in industry~\parencite{Korte2018}, the large number of possible solutions and the complex constraints often involved pose significant computational challenges. Even state-of-the-art classical metaheuristics struggle to find high-quality solutions for large instances of well-known combinatorial problems such as the Knapsack Problem~\parencite{Kellerer2004} and the Traveling Salesman Problem~\parencite{Punnen2007}.

Among classical methods, Simulated Annealing (SA) is a widely adopted metaheuristic for addressing combinatorial problems~\parencite{Kirkpatrick1983}. In SA, a cost function assigns an energy value to each possible solution, thereby defining an energy landscape. The algorithm explores this landscape through probabilistic transitions to new candidate solutions, with acceptance probabilities controlled by a computational temperature parameter. At high temperatures, uphill moves are more frequently accepted, allowing the algorithm to escape local minima. As the temperature gradually decreases, the acceptance of such moves becomes less likely, guiding the search to focus on promising regions and converge toward high-quality solutions.

In contrast, Quantum Annealing does not rely on thermal fluctuations. Instead, it exploits the adiabatic theorem of quantum mechanics to evolve a quantum system from the ground state of an Initial Hamiltonian $H_I$ to the ground state of a Problem Hamiltonian $H_P$~\parencite{Lucas2014} and makes use of quantum tunneling~\parencite{Kadowaki1998}, which helps the system to transit among the different potential minima. This final Hamiltonian encodes the function to be minimized, and its ground state corresponds to the optimal solution. Since both methods can be adjusted to explore the same optimization landscape, SA serves as a natural benchmark for evaluating the performance of QA. The following subsections provide a more detailed overview of these two metaheuristics and D-Wave's hybrid solver, which combines classical and quantum methods to address larger problem instances beyond the capacity of current quantum hardware.

\subsection{Simulated Annealing}

Simulated Annealing is a classical optimization algorithm inspired by the annealing process in metallurgy, where a material is heated to a high temperature and then gradually cooled to reduce structural defects, allowing it to reach a stable, low-energy crystalline configuration~\parencite{Kirkpatrick1983}. Simulated Annealing requires a function to be minimized to simulate this process in the context of optimization, allowing the problem's solution space to be viewed as a landscape of configurations, each associated with a cost (or energy) value defined by a Problem Hamiltonian $H_P$. A computational, non-physical temperature parameter controls the probability of transitioning between solutions. Through a controlled cooling schedule, SA gradually converges to optimal or near-optimal solutions, analogous to a crystalline solid achieving its lowest-energy state~\parencite{Henderson2003}. 

The algorithm starts by setting a high temperature parameter $T$ and selecting a random initial solution $\xi$. At each iteration, a new solution $\xi'$
in the neighborhood of $\xi$ is generated using a neighborhood operator~\parencite{Siddique2016}. The algorithm then decides whether to accept $\xi'$  as the current solution based on the transition probability
\begin{equation}
    p = \min \left\{ 1, \exp\!\left(-\beta \, \Delta H\right) \right\}, 
    \qquad \Delta H = H(\xi') - H(\xi),
    \label{eq:prob_trans}
\end{equation}

\noindent where $\beta = 1/T$ is the inverse temperature. At high temperatures, $\exp{(-\beta \Delta H)} \approx 1$, meaning that transitions to worse quality solutions are likely to be accepted, allowing the algorithm to escape local minima early in the search. As $T$ decreases, the algorithm becomes more selective, increasingly rejecting
worse solutions. Because improvements to better solutions are always accepted, SA can effectively refine the search around promising regions in the solution space.

While SA is considered an efficient classical approach to optimization problems, it also has several limitations. Because cooling the system to absolute zero is impractical, the algorithm can only approximate the ideal annealing process~\parencite{Nikolaev2010}. In large solution spaces, this may cause the algorithm to become trapped in local minima. Slower cooling schedules allow for more extensive exploration before convergence and can yield better solutions but significantly increase runtime. Careful parameter tuning for the problem of interest -- such as selecting a promising initial solution, adjusting the number of iterations per temperature (i.e., the length of the Markov chains), and elaborating neighborhood operators and temperature schedules that are well adapted to the problem -- can help balance exploration with convergence but will not generalize across all instances of the same problem~\parencite{Ledesma2008}. Furthermore, since SA is a stochastic method, multiple independent runs are typically performed to increase the likelihood of findi
ng a near-optimal or global optimal solution.

\subsection{Gurobi Optimizer}

As a classical, exact optimization baseline, we employ the commercial solver Gurobi Optimizer (\textcite{Gurobi}), a state-of-the-art mixed-integer programming solver widely used in operations research and industry. Although Gurobi implements branch-and-bound and cutting-plane methods that can, in principle, achieve optimality, in this work, we use it with practical stopping criteria based on time limits, in accordance with the settings adopted for the other methods. Heuristic components were kept enabled to accelerate the identification of good feasible solutions, which may introduce some variability across runs, but does not alter the solver’s fundamental character. After this initial introduction, we will refer to it simply as Gurobi.

\subsection{D-Wave's Quantum Annealing and Hybrid Solvers}

Quantum Annealing is a quantum metaheuristic based on the adiabatic evolution of a quantum system toward the minimum of a cost function~\parencite{McGeoch2014}. In recent years, quantum annealers have been applied to small and intermediate-scale problems, enabling proof-of-concept in many operational problems, such as scheduling~\parencite{Rieffel2014, Yu2021}, traffic flow~\parencite{Neukart2017, Inoue2021}, and vehicle routing~\parencite{Feld2019, Irie2019}.

A practical implementation of QA is provided by D-Wave Systems, whose quantum processors can address problems expressed as \textit{Quadratic Unconstrained Binary Optimization} (QUBO) Hamiltonians~\parencite{Glover2022}. In this class of Hamiltonians, the cost function to be minimized must be written in the form
\begin{equation}
H_P = \sum_i h_i x_i + \sum_{i>j} J_{ij} x_i x_j,
\label{eq:qubo_main}
\end{equation}
\noindent where $x_i$ and $x_j$ are binary variables, and $h_i$ and $J_{ij}$ are real-valued coefficients that can be adjusted to encode the parameters of the optimization task. To address a given problem, Eq.~(\ref{eq:qubo_main}) must be formulated such that lower-energy states correspond to better solutions, with the ground state representing the optimal one. Since QUBO models are unconstrained, not only the objective function but also all the feasibility conditions (i.e., the set of constraints) of the task must also be embedded in Eq.~{(\ref{eq:qubo_main})} as penalty terms. This modeling requirement distinguishes QA from classical metaheuristics, which can enforce constraints through procedural logic (e.g., conditional statements to discard infeasible solutions).

In QA, the quantum system is initially prepared in the ground state of an Initial Hamiltonian $H_I$ and then slowly evolved into the Problem Hamiltonian $H_P$. According to the adiabatic theorem, if the evolution is sufficiently slow and the system remains isolated, it will remain in the ground state with probability close to 1, reaching the optimal solution encoded in $H_P$ at the end of the interpolation~\parencite{Albash2018}. In practice, however, current quantum annealers are limited by environmental noise, control precision, and qubit connectivity, which restrict the problem sizes they can handle~\parencite{Rajak2022}. Despite these limitations, QA offers a potential advantage over classical metaheuristics by exploiting \textit{quantum tunneling} – a quantum phenomenon that allows the system to transition through high but narrow energy barriers in the energy landscape, rather than having to climb over them as in classical methods~\parencite{Kadowaki1998}. This mechanism can potentially provide an advantage in combinatorial optimization by enabling exploration of rugged landscapes where classical approaches are prone to getting trapped in local minima. For readers interested in the quantum computing aspects of this optimization method, we refer to the works of \textcite{Albash2018} and \textcite{Yarkoni2022}.

To overcome the input-size limitations of Quantum Annealing hardware, D-Wave introduced the Hybrid Solver Service (HSS)~\parencite{dwavetechupdate}, a portfolio of hybrid solvers that combine Quantum Annealing with classical metaheuristics to address problem instances much larger than those that can be directly embedded in current-generation quantum annealers~\parencite{dwavereport}. The most general of these solvers, the Constrained Quadratic Model (CQM) Solver supports binary, integer and real variables, while allowing the explicit inclusion of linear and quadratic constraints~\parencite{DWAVE2022}. The solver allows users to formulate optimization tasks in a higher-level representation that is internally reduced into the QUBO formulation used by both the quantum component of the hybrid workflow~\parencite{dwavereport}. 

In the HSS architecture, each hybrid solver comprises a classical module and a quantum module operating in tandem~\parencite{dwavereport}. The classical component executes in parallel a set of classical metaheuristics to explore the solution space and search for high-quality solutions, while the quantum module formulates quantum queries — subproblems small enough to be solved directly on a D-Wave quantum annealer~\parencite{dwavetechupdate}. These reduced  QUBO instances are sent to D-Wave’s quantum annealers for multiple runs. The results from these quantum queries are then incorporated back into the classical workflow to refine the current pool of candidate solutions or to guide the search toward more promising regions of the solution space \parencite{dwavereport}. Since both QA and the integrated classical methods are stochastic, the overall output of the hybrid solver is also probabilistic, and multiple runs are typically performed to increase the chances of obtaining better-quality solutions.

According to D-Wave’s internal benchmarks~\parencite{dwavetechupdate}, hybrid solvers in the HSS outperformed 37 state-of-the-art classical solvers from the public MQLib repository in most of the tested instances. As explained by D-Wave~\parencite{dwavetechupdate, DWAVE2022}, internal versions of the hybrid solvers can operate in two distinct modes. In the \emph{hybrid workflow}, the quantum module formulates quantum queries, sends them to the QPU, receives replies, and incorporates them back into the classical search process. In the \emph{heuristic workflow}, the same solver runs with quantum queries disabled, relying solely on classical heuristics. Internal tests comparing performance with and without quantum queries enabled showed consistently faster convergence toward better-quality solutions when the quantum module was active~\parencite{dwavetechupdate}. D-Wave referred to this effect as \emph{quantum acceleration of classical heuristics}~\parencite{dwavetechupdate} or \emph{hybrid acceleration}~\parencite{DWAVE2022}, in which quantum queries guide the classical solver to explore more promising regions of the search space and reach improved solutions faster than would otherwise be possible~\parencite{dwavetechupdate, DWAVE2022}. Although the expression \emph{quantum (hybrid) acceleration} no longer appears in the most recent technical reports, the underlying description of the mechanism — where quantum responses are used to guide the heuristic search or improve the quality of the solution pool — is still maintained~\parencite{dwavereport}.

It is important to emphasize that the internal mechanisms of the HSS are not accessible to users. The relative contributions of its quantum and classical components, the frequency of quantum queries, and the criteria governing their invocation are not publicly disclosed. Furthermore, the version of the Hybrid Solver Service available to users does not permit disabling the quantum module \parencite{DWAVE2022} and accessing the \emph{heuristic workflow}, making it impossible to replicate the hybrid-versus-heuristic workflows comparison described in D-Wave’s internal reports. Therefore, we regard the solver as a black-box hybrid optimizer that integrates classical heuristics running in parallel with calls to Quantum Annealing. In the following, our analysis focuses on its empirical performance and the results obtained, without making quantitative claims about the magnitude of the quantum contribution.

\section{Strategy Formulation for Optimizing Inventory Allocation in Gravity-Flow Racks}
\label{SecModel}

In warehouses equipped with gravity-flow racks, item-picking operations can lead to reinsertions. While a certain number of these operations are expected in FIFO-based systems, high quantities can compromise warehouse efficiency. In particular, if items frequently requested together end up scattered across multiple shelves or placed behind rarely requested items, future picking operations can create a cycle of new reinsertions that otherwise would be unnecessary.

To address this issue, we propose a strategy for optimizing the simultaneous allocation of multiple items awaiting distribution across the available shelves. These items may include newly arrived items, items removed for access and requiring reinsertion, or a combination of both. Our approach prioritizes the joint allocation (i.e., placement on the same shelf) of items that are frequently requested together by production lines. Using a machine learning protocol, these preferences are captured through matching parameters derived from historical demand data. The following section describes how this strategy is encoded in an optimization model.

\subsection{Formulation of the Strategy as an Optimization Problem}\label{SecForm}

Consider a scenario where $N$ distinct items must be distributed across $M$ gravity-flow shelves. For each shelf $m$ ($m = 1, \ldots, M$), let $R_m$ denote its remaining capacity limit, which may represent maximum allowable weight, length, or number of items. For each item $\alpha$ ($\alpha=1, \ldots, N$), we associated a value $v_{\alpha}$, expressed with the same units of $R_m$. The remaining capacity limit thus imposes an upper bound on the total possible allocations per shelf. Hence, on each shelf, allocations must satisfy the constraint
\begin{equation}\label{inequality}
\sum_{\alpha=1}^{N} v_{\alpha} x_{\alpha}^m \leq R_m,
\end{equation}
\noindent where $x_{\alpha}^m$ is a binary variable that equals 1 if item $\alpha$ is assigned to shelf $m$, and 0 otherwise.

To capture how frequently items are requested together in the production line’s demand history, we employ a machine learning protocol that generates matching parameters $0 \leq \lambda_{\alpha \beta } \leq 1$ quantifying the preference for assigning items $\alpha$ and $\beta$ to the same shelf, with smaller values of $\lambda_{\alpha \beta }$ indicating a stronger priority for joint allocation. Consequently, for $\alpha \neq \beta$, three scenarios arise: (i) $\lambda_{\alpha \beta } = 0$, indicating a strong preference for assigning items $\alpha$ and $\beta$ to the same shelf, (ii) $\lambda_{\alpha \beta } = 1$, indicating a strong preference for assigning them to different shelves, and (iii) intermediate values, reflecting the priority of these items being allocated together compared to other pairs.

In real-world warehouses, each shelf $m$ may already contain $P_m$ previously allocated items. To account for preferences between new items and pre-allocated items, we introduce additional matching parameters $\lambda_{\alpha\tau}^{(m)}$ for ($\tau = 1, \ldots, P_m$), which quantify the priority of allocating item $\alpha$ to shelf 
$m$ relative to the existing items. Thus, if item $\alpha$ is placed on shelf $m$, the sum of all the interactions between $\alpha$ and the pre-allocated items is equal to $\sum_{\tau = 1}^{P_m} \lambda_{\alpha \tau}^{(m)}$. 

We interpret the matching parameters as cost values to optimize the distribution of new items across the available shelves, prioritizing the joint allocation of items frequently requested together and thereby reducing the likelihood of future reinsertions. This way, the cost of assigning new items to a single shelf $m$ consists of two terms: the sum of matching parameters between pairs of newly allocated items ($\lambda_{\alpha \beta}$) and the sum of matching parameters between each new item and the items already present on that shelf ($\lambda_{\alpha \tau}^{(m)}$). This cost is expressed as
\begin{equation}\label{eq3}
    \Lambda_m = \sum_ {\substack{\alpha, \beta=1 \\ (\alpha > \beta)}}^N  \lambda_{\alpha \beta}    x_{\alpha}^{m} x_{\beta}^{m} +  \sum_{\alpha=1}^N \sum_{\tau=1}^{P_m}  
   \lambda_{\alpha \tau}^{(m)}  x_{\alpha}^m.
\end{equation}
\noindent 
The objective is to minimize the total cost across all shelves, thereby promoting allocations in which items frequently requested together are placed on the same shelf. Formally, the objective function is given by
\begin{equation}\label{função_objetivo}
\Lambda = \sum_{m=1}^{M} \Lambda_m,
\end{equation}
subject to the conditions that (i) each of the $N$ items must be allocated to exactly one shelf, and (ii) the capacity constraints defined in Eq.~(\ref{inequality}) must be satisfied.

\subsection{Complexity of the problem}

The computational complexity of the proposed problem arises from its connection to the Knapsack problem family~\parencite{Cacchiani2022a}. Specifically, Eq.~(\ref{eq3}) can be viewed as a modified version of the Required Multiple Quadratic Knapsack Problem (RMQKP)~\parencite{Cacchiani2022b}. The modification lies in the inclusion of a term accounting for items already present in the knapsacks (referred to as shelves in our context). As the number of items and shelves increases, the possible configurations grows exponentially, resulting in similar NP (Non-Deterministic Polynomial Time) behavior observed in RMQKP~\parencite{Cacchiani2022b}. The NP class of problems is characterized by the intractability of their solution in polynomial time, yet the correctness of a given solution can be verified in polynomial time~\parencite{Sipser2006}. This class incorporates various industrially relevant problems, including the Traveling Salesman Problem and the aforementioned Knapsack Problem. A more detailed discussion on the scaling of the solution space size for the addressed problem, including the relevant constraints, is provided in the Supplementary Material, where we derive a lower bound for the number of possible solutions.

\subsection{Problem Hamiltonian}\label{SecHamiltonian}

As discussed in Sec.~\ref{SecMethods}, both Simulated Annealing and Quantum Annealing, along with D-Wave's hybrid solvers, rely on a Hamiltonian that encodes the possible solutions of the optimization task and thereby defines the associated energy landscape. In this formulation, each candidate solution is assigned an energy value, with lower energies corresponding to better solutions. For implementation on D-Wave’s quantum annealers and hybrid solvers, the Hamiltonian must be expressed in QUBO form.

To encode the proposed strategy for the simultaneous allocation of $N$ items across $M$ shelves, while incorporating all relevant constraints, we define a Problem Hamiltonian of the form
$H_P = H_A + H_B + H_C$, where
\begin{subequations}\label{ProblemHamiltonian}
    \begin{equation}\label{Hc}
        H_A = A \sum_{\alpha=1}^{N}\left(1 - \sum_{m=1}^{M} x_{\alpha}^m\right)^2,
    \end{equation}
    \begin{equation}\label{Hd}
        H_B = B \sum_{m=1}^{M} \left(\sum_ {\substack{\alpha, \beta=1 \\ (\alpha > \beta)}}^N  \lambda_{\alpha \beta}    x_{\alpha}^{m} x_{\beta}^{m} +  \sum_{\alpha=1}^N \sum_{\tau=1}^{P_m}  
        \lambda_{\alpha \tau}^{(m)}  x_{\alpha}^m \right),
    \end{equation}
    \begin{equation}\label{He}
        H_C = C \sum_{m=1}^M \left(\sum_{\alpha=1}^{N} v_{\alpha}  x_{\alpha}^{m} + \braket{\bm{2} |\bm{a}_m} - R_m\right)^2.
    \end{equation}
\end{subequations}

\noindent Each term in $H_P$ enforces a requirement of the allocation task. The coefficients $A$, $B$, and $C$ act as penalty parameters that balance constraint satisfaction and objective minimization. $H_A$ penalizes solutions in which any item $\alpha$ is either allocated to multiple shelves or none, thereby enforcing that each item is assigned to exactly one shelf. $H_B$ follows from Eq.~(\ref{função_objetivo}) and represents the cost $\Lambda$ to be minimized, favoring allocations in which items frequently requested together are assigned to the same shelves. Finally, $H_C$ penalizes solutions that violate the capacity constraints in Eq.~(\ref{inequality}). The term $\braket{\bm{2} |\bm{a}_m} = \sum_{l = 0}^{\Omega_m} 2^l a_l^{m}$, with $\Omega_m = \lfloor {\log_2{R_m}} \rfloor$, is a binary expansion over the slack variables $a_l^m$, which are introduced to transform the capacity constraints into equalities – a common technique to make optimization models that use inequalities compatible with QUBO formulations. In our formulation, these slack variables act as virtual items with value $v_{\alpha} = 2^l$, representing the unused portion of the shelf capacity, allowing the inequality to be rewritten as an equality in QUBO form while ensuring that shelf limits are respected. Together, these terms ensure that every item is uniquely assigned, capacity limits are respected, and the overall allocation cost is minimized.

The parameters $A$, $B$, and $C$ must be selected to ensure that constraint violations are correctly penalized. This can be achieved by setting $A$ and $C$ significantly larger than $B$. Additional adjustments may be needed to obtain high-quality solutions depending on the metaheuristic used to apply the strategy. Since $H_P$ is formulated as a QUBO, it can be implemented in D-Wave's quantum annealers and hybrid solvers. Moreover, the same formulation is compatible with other quantum metaheuristics not part of this work, such as QAOA~\parencite{Farhi2014}. An implementation of our strategy using QAOA in a trapped-ion hardware is reported by \textcite{Ricardo2024}.

\section{Results and Discussions}\label{SecResults}

In this section, we apply our strategy to various warehouse configurations and compare the results obtained using SA, Gurobi and D-Wave's CQM Solver, utilizing the Hamiltonian introduced in Eqs.~(\ref{Hc}), (\ref{Hd}), and (\ref{He}). We begin by outlining the implementation details of each method, including the neighborhood operators used in the two different implementations of Simulated Annealing. We then compare (i) the quality of the results obtained by the methods within a fixed time, evaluating their performance across warehouses of varying sizes and different numbers of items to be allocated, and (ii) the time required for the methods to achieve solutions of comparable quality within a specified error margin. For all tests in this section, we set $v_{\alpha}=1$ for every item $\alpha$, so that shelf capacities are counted in unit loads (number of items/pallets) rather than weight or volume. This choice simplifies the analysis but does not limit the generality of the formulation: alternative definitions of $v_{\alpha}$ could incorporate attributes such as weight or volume, and the framework naturally admits multiobjective extensions in which several constraints or performance criteria are optimized simultaneously.

\subsection{Details of Metaheuristic Implementations}

\textbf{Simulated Annealing:} As previously discussed in Sec.~\ref{SecMethods}, several choices are made in the implementation of SA. One critical choice is the neighborhood operator, a function that takes a solution $\xi$ as input and produces a neighboring solution $\xi'$. In the simulations, we implement two variants of SA, distinguished by the representation of the problem and by the operators employed to explore the solution space. 

The first variant, denoted RS-SA, employs a combination of two operators that are specifically adapted to the warehouse allocation problem. The two operators considered are the Real operator, which moves an item from one shelf to another, and the Swap operator, which exchanges two items between their shelves~\parencite{Chen2016}. At each iteration, one of these operators is chosen with equal probability. Since these operators always generate feasible solutions for the Hamiltonian $H_A$, RS-SA directly enforces the single-assignment constraint during the search process.

A second variant, denoted INT-SA, uses an integer encoding of the problem, where a single decision variable per item indicates the shelf to which it is allocated. The Real and Swap operators are again employed as neighborhood operators, acting directly on the integer variables. While this representation is fully compatible with SA and allows for a more direct modeling of the allocation task, it cannot be implemented on D-Wave’s CQM hybrid solver because it relies on Kronecker delta functions that cannot be expressed in the QUBO formalism. For this reason, INT-SA is, for now, restricted to classical computation, providing an alternative formulation that highlights the flexibility of SA beyond binary encoding.

For the RS-SA and INT-SA implementations, the single-assignment constraint is intrinsically guaranteed by the neighborhood operators, so no penalty parameter $A$ was required. In these cases, $C$ was set to a sufficiently large value relative to $B$ to ensure that shelf capacity constraints were never violated, while keeping the objective term properly balanced.

\textbf{Gurobi Optimizer:}  As a solver-based baseline, we formulated the allocation task as a mixed-integer quadratic program (MIQP) and solved it with the Gurobi Optimizer \parencite{Gurobi}. The model is equivalent to the QUBO Hamiltonian used in the metaheuristic and hybrid approaches, with binary variables $x_{i,m}$ indicating whether item $i$ is assigned to shelf $m$, subject to unique-assignment and capacity constraints. We adopted runtime as the stopping criterion rather than requiring certified optimality. To guide the solver, presolve routines were activated at their most aggressive level to accelerate convergence, and Gurobi’s heuristic module was set to its maximum, reflecting our focus on quickly identifying high-quality feasible solutions rather than on certifying optimality. Under these settings, the solver produced feasible solutions of good quality within the allotted time, while also reporting lower bounds that allowed us to track the residual gap. Because the use of heuristics may introduce variability across executions, we performed multiple independent runs and report statistics accordingly.

\textbf{CQM Solver:} In D-Wave's hybrid method, a user-defined time parameter sets a limit on the running time of the algorithm~\parencite{dwavereport}.  It specifies the maximum duration for which the hybrid solver will operate. For small instances, there is a minimal time requirement of 5 seconds to ensure the algorithm has enough time to explore the solution space~\parencite{SolverProp2025}. For larger problems, the required time increases accordingly. By carefully setting this parameter, one can balance computational efficiency and solution quality, ensuring satisfactory results are achieved within the specified time. In our simulations, the time limit was set to the minimum required, reflecting the need for warehouse instructions to be issued with minimal delay. However, the appropriate duration may vary according to the operational requirements of each case. 

As for the constraints, the hybrid solver allows them to be specified either as \textit{soft} or \textit{hard}. In the former case, the associated penalty weights are automatically balanced within the specified runtime, and their exact values are not user-controllable. In the latter case, the solver guarantees feasibility by strictly enforcing the constraints. In this work, we adopted hard constraints to ensure that all solutions considered during the tests were feasible.

\subsection{Results and Comparative Analysis}

We begin by comparing the solutions obtained by the CQM Solver, two variants of Simulated Annealing (RS-SA and INT-SA), and the Gurobi Optimizer across benchmark warehouse configurations of increasing sizes. For all configurations, we adopt a square, simplified layout in which the number of shelves is equal to the number of positions per shelf, labeling each warehouse configuration as $M\times M$. Although such a layout is not typical in real-world warehouses, it serves as an effective basis for evaluating the metaheuristics' performance as the problem scales up. 

In each $M \times M $ warehouse configuration generated as an instance for the task, 20\% of the available positions are initially filled with items, uniformly distributed across the shelves. We then address a series of independent allocation tasks in which new items corresponding to 10\%, 20\%, 30\%, 40\%, 50\%, and 60\% of the total warehouse capacity must be inserted, resulting in final occupancy levels ranging from 30\% to 80\%.  For illustrative purposes, in a $20 \times 20$ warehouse (400 total positions), considering that there are 80 items (20\%) previously distributed on the shelves (four per shelf), we first consider the task of allocating 40 additional items (10\%), resulting in a final warehouse occupancy of 30\%. Subsequent tests follow this pattern until the final task, in which 240 additional items (60\%) are inserted, reaching a final warehouse occupancy of 80\%. 

\begin{figure*}
    \centering
\includegraphics[width=1\linewidth]{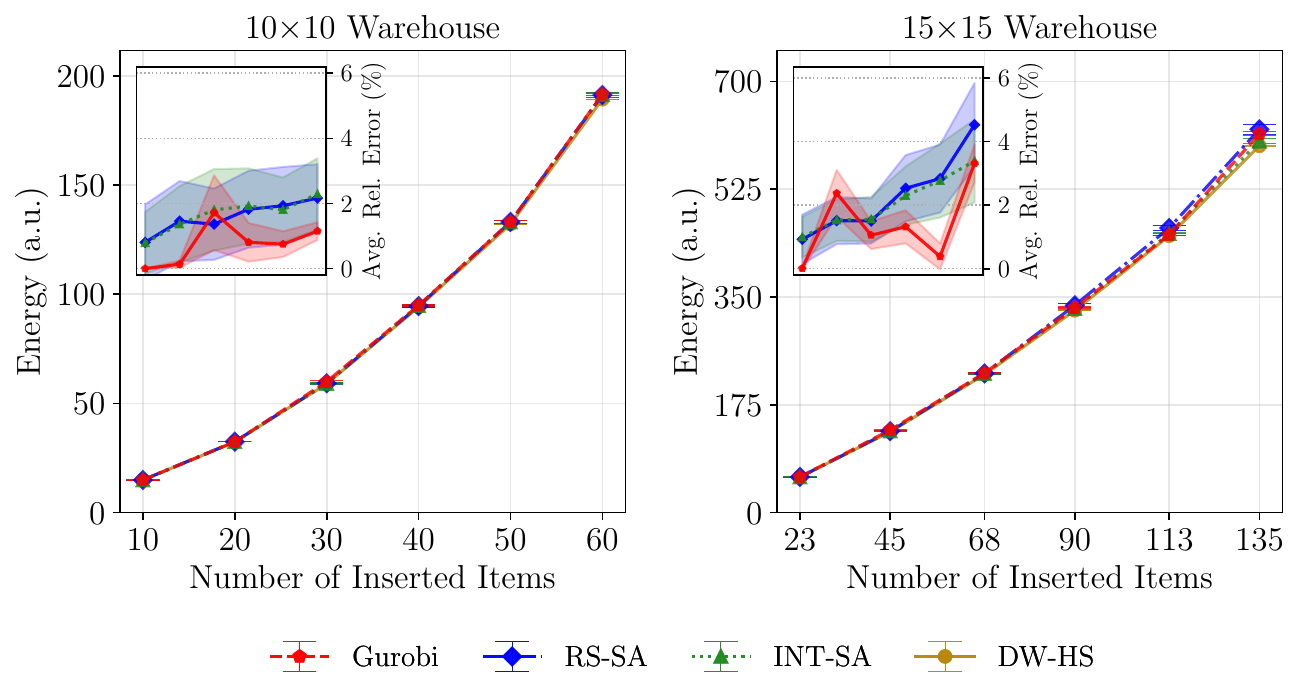}
\caption{Average energy obtained by the CQM Solver, the Gurobi Optimizer, and the two Simulated Annealing variants (RS-SA and INT-SA) for the $10 \times 10$ and $15 \times 15$ warehouse configurations, shown in panels (a) and (b), respectively. Error bars represent the standard deviation across 100 runs for each SA variant and Gurobi, and 50 runs for the CQM Solver. The inset reports the relative error of each method with respect to the average energy achieved by the CQM Solver. The time limit for each method was set to 5 seconds, corresponding to the minimum runtime required by the CQM Solver across all instances.}
\label{energiasTempoFixoa}
\end{figure*}

\begin{figure*}[h]
    \centering
\includegraphics[width=1\linewidth]{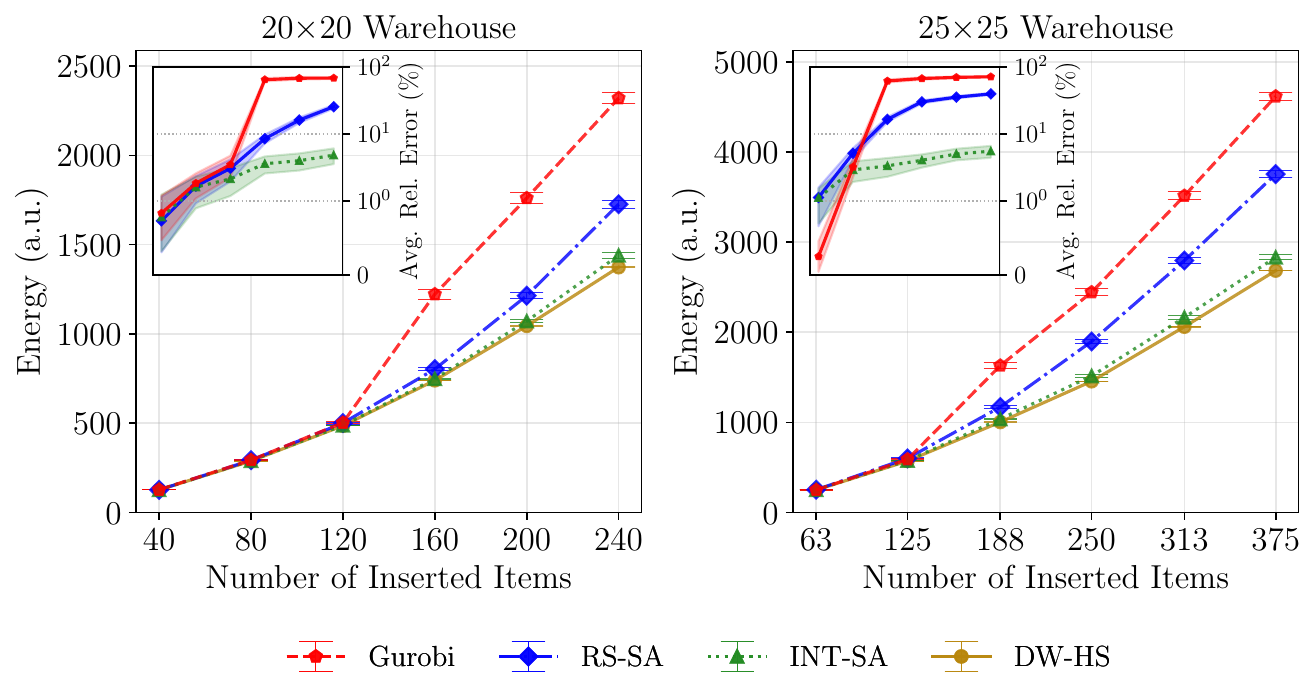}
\caption{Average energy obtained by the CQM Solver, the Gurobi Optimizer, and the two Simulated Annealing variants (RS-SA and INT-SA) for the $20 \times 20$ and $25 \times 25$ warehouse configurations, shown in panels (a) and (b), respectively. Error bars represent the standard deviation across 100 runs for each SA variant and Gurobi, and 50 runs for the CQM Solver. The inset reports the relative error of each method with respect to the average energy achieved by the CQM Solver. The time limit for each SA variant was set to 5 seconds, equal to the minimum runtime required by the CQM Solver for all instances, except for the task of allocating  312 and 375 items in the $25\times25$ configuration, which required 7.04 and 9.79 seconds, respectively.}
\label{energiasTempoFixoc}
\end{figure*}

All simulations were performed under identical time constraints, with the time limit for each SA variant and for Gurobi set equal to the minimum runtime required by the CQM Solver. We present results for four warehouse configurations grouped into two figures: the $10 \times 10$ and $15 \times 15$ configurations (Fig.~\ref{energiasTempoFixoa}), and the $20 \times 20$ and $25 \times 25$ configurations (Fig.~\ref{energiasTempoFixoc}). In each figure, the main plots display the average energy obtained from 100 runs of each SA implementation, 100 runs of the Gurobi Optimizer, and 50 runs from the CQM Solver, with error bars indicating the corresponding standard deviations. Each plot includes an inset displaying the average relative error of each method with respect to the mean energy obtained by the CQM Solver, which consistently outperformed the other implementations and reached the best known solution in every instance. Execution times are reported in the respective figure captions. 

For the RS-SA and INT-SA implementations, to generate the data points presented in the main plots, we tested two Markov chain lengths for each allocation task, set to $N$ and $N \times M$. For each chain, 100 independent runs were executed for both SA variants. For every data point, the chain producing the lowest average energy was selected, which means that different points within the same plot may correspond to different chain lengths. These two chain lengths were chosen based on preliminary tests, where they consistently outperformed other alternatives across multiple instances. All simulations were executed on an Intel Core i7-6700K CPU (4.00 GHz, 8 cores) with 16 GB of RAM. All datasets used to produce the figures are publicly available for reproducibility~\parencite{Valerio2025}.

An analysis of Fig.~\ref{energiasTempoFixoa} and Fig.~\ref{energiasTempoFixoc} shows that the Gurobi Optimizer performs competitively in smaller instances, producing results close to those of the D-Wave hybrid solver and both Simulated Annealing variants under the same time limits. This is particularly evident in Figs.~\SubFig{energiasTempoFixoa}{(a)} and ~\SubFig{energiasTempoFixoa}{(b)}, where Gurobi maintains a relative error below 4\% with respect to the CQM Solver across all insertion levels. As the problem size grows, however, this relative performance progressively weakens. In Fig.~\SubFig{energiasTempoFixoc}{(a)}, Gurobi remains close to the SA variants up to the 30\% insertion level (with relative errors below $4\%$). Beyond this point, however, its performance deteriorates rapidly, with deviations exceeding $50\%$ relative to RS-SA and more than $60\%$ relative to INT-SA. A similar trend is observed in Fig.~\SubFig{energiasTempoFixoc}{(b)}, where Gurobi stays competitive only up to the $20\%$ insertion level. From 30\% onward, its relative error surpasses about 25 -- 40\% compared to RS-SA and more than 60\% compared to INT-SA, highlighting a marked loss of solution quality.

In contrast, the comparison between the two SA variants is more straightforward. Across all warehouse configurations, INT-SA consistently outperforms RS-SA, exhibiting relative errors never exceeding 6\% with respect to the CQM Solver. RS-SA, in turn, remains below 5\% in smaller instances but progressively deteriorates as the problem size grows, reaching a maximum relative error of approximately 25\% in the 20$\times$20 warehouse and 40\% in the 25$\times$25 configuration. This systematic difference indicates the advantage of adopting an integer formulation combined with problem-specific neighborhood operators, which together enhance the search efficiency of INT-SA. Overall, however, the D-Wave hybrid solver remains superior to all classical heuristics and to Gurobi, with the performance gap widening as both the number of shelves and the insertion level increase.

To investigate the performance differences further, we analyzed the time required for RS-SA, INT-SA and the Gurobi Optimizer to achieve results comparable to those of the CQM Solver. As shown in the previous analysis, the CQM Solver consistently obtains higher-quality solutions, so we set the average of these results as the benchmark for both the metaheuristics and Gurobi. Since expecting these methods to exactly match the hybrid solver would be unrealistic, we defined solutions within 3\% of the benchmark as being of comparable quality, considering that the maximum percentage standard deviation of the hybrid solver was only $0.14\%$ (in the 50\% insertion task with 313 items). The corresponding runtimes are reported in Table~\ref{tab:time_comparison}, which shows that INT-SA requires substantially less time than RS-SA to reach acceptable solutions, while Gurobi, on average, lies between the two SA variants in terms of runtime. However, its results exhibit substantially larger standard deviations, indicating less stability across independent runs. Nevertheless, both metaheuristics and Gurobi require significantly longer times than the CQM Solver, underscoring the D-Wave hybrid approach's efficiency in handling larger warehouse configurations.

\begin{table}[h]
\centering
\begin{tabular}{|c|c|c|c|c|}
\hline
\textbf{Items (\%)} & \textbf{DW-HS} & \textbf{RS-SA} & \textbf{INT-SA} & \textbf{Gurobi} \\ \hline
& $t_{\text{min}}$ (s) & $t_{\text{mean}} \pm \Delta t$ (s) & $t_{\text{mean}} \pm \Delta t$ (s) & $t_{\text{mean}} \pm \Delta t$ (s) \\ \hline
63 (10\%)  & 5.00  & 5.21 $\pm$ 1.06     & 0.00 $\pm$ 0.00 & 0.48 $\pm$ 0.15\\ \hline
125 (20\%) & 5.00  & 38.05 $\pm$ 4.85    & 1.70 $\pm$ 0.46 & 12.52 $\pm$ 1.59 \\ \hline
188 (30\%) & 5.00  & 122.2 $\pm$ 13.47   & 6.10 $\pm$ 0.79 & 54.84 $\pm$ 20.69 \\ \hline
250 (40\%) & 5.00  & 271.22 $\pm$ 32.15  & 12.71 $\pm$ 1.55 & 253.46 $\pm$ 210.01 \\ \hline
313 (50\%) & 7.04  & 492.88 $\pm$ 57.24  & 23.79 $\pm$ 4.31 & 561.56 $\pm$ 501.66 \\ \hline
375 (60\%) & 9.79  & 849.83 $\pm$ 133.12 & 40.20 $\pm$ 10.10 & $1080.44 \pm 973.68$ \\ \hline
\end{tabular}
\caption{Runtime comparison of the DW-HS (CQM Solver), RS-SA, INT-SA, and Gurobi for the $25 \times 25$ warehouse configuration across different insertion levels. The second column shows the minimum runtime of the hybrid solver. The third and fourth columns report the average runtime and standard deviation for RS-SA and INT-SA, respectively, to reach solutions within a 3\% margin of the CQM Solver’s average.}
\label{tab:time_comparison}
\end{table}

To explore the scalability limits of our implementations, we extended the tests to a larger $30 \times 30$ warehouse configuration, maintaining the same pattern of independent allocation tasks in which items were inserted in 10\% increments of the total warehouse capacity (i.e., from 90 to 540 items). Table~\ref{tab:rel_errors_dwave} reports the relative error of RS-SA, INT-SA, Gurobi, and the CQM Solver (DW-HS), all evaluated under fixed runtimes equal to the minimum required by the CQM Solver at each insertion level. The corresponding execution times are indicated in the table caption. All relative errors are computed with respect to the best solution found by the CQM Solver. Under this criterion, RS-SA showed increasingly large deviations as the insertion level grew, whereas the CQM Solver consistently achieved the best solutions across all cases. INT-SA remained comparatively accurate, with deviations below 10\% throughout, reaching a maximum of 9.8\% at the 60\% insertion level. In contrast, Gurobi exhibited a substantial deterioration in solution quality, with relative errors surpassing those of RS-SA from the 20\% insertion level onward and exceeding 100\% in the largest allocation tasks, underscoring its limitations under the imposed runtime constraints.

Following this analysis, we once again investigated the runtime required for INT-SA to reach solutions within a 3\% margin of the CQM Solver’s average. In these simulations, even the larger Markov chain (with length $N \times M$) began to show limitations. For the 10\%, 20\%, and 30\% allocation tasks, we were still able to obtain 100 valid runs (within a 3\% margin of the CQM Solver’s average), by performing 150 runs per task and retaining the first 100 that converged successfully. However, starting at the 40\% insertion level, only 21 out of 150 runs met the error threshold using the $N \times M$ chain length. To reach 100 valid samples, it was necessary to significantly increase the chain length, which raised the average runtime to approximately 300 seconds. For the subsequent 50\% and 60\% insertion tasks, no tested Markov chain was able to produce 100 valid runs within 3\% error in under 10 minutes per run. The amount of testing required to attempt convergence in these cases was so extensive that it became impractical to define a representative execution time. These findings suggest that we reached the practical computational limits of our hardware setup. Similarly, the Gurobi Optimizer required on average $6.20 \pm 2.46$ minutes for the 40\% insertion task, $21.55 \pm 13.36$ minutes for the 50\% task, and $50.34 \pm 11.57$ minutes for the 60\% task, indicating that, on the implemented hardware, both INT-SA and Gurobi became impractical beyond the 40\% insertion level.

\begin{table}[htbp]
\centering
\caption{Average relative errors (R.E.) of DW-HS, RS-SA, INT-SA, and the Gurobi Optimizer with respect to the best solution found by DW-HS, for each insertion level in the $30 \times 30$ warehouse configuration. The corresponding minimum runtimes required by the D-Wave hybrid solver are: 10\% -- 5.00s, 20\% -- 5.00s, 30\% -- 6.48s, 40\% -- 10.91s, 50\% -- 16.47s, and 60\% -- 23.17s.}
\label{tab:rel_errors_dwave}
\begin{tabular}{ccccc}
\hline
\textbf{$N$} & \textbf{DW-HS R.E. (\%)} & \textbf{RS-SA R.E. (\%)} & \textbf{INT-SA R.E. (\%)} & \textbf{Gurobi R.E. (\%)} \\
\hline
90 & 0.00 $\pm$ 0.00 & 2.20 $\pm$ 0.70 & 0.43 $\pm$ 0.25 & 0.05 $\pm$ 0.08 \\
180 & 0.11 $\pm$ 0.04 & 24.96 $\pm$ 2.06 & 2.54 $\pm$ 0.86 & 87.3 $\pm$ 3.16 \\
270 & 0.26 $\pm$ 0.14 & 55.14 $\pm$ 2.60 & 6.88 $\pm$ 1.66 & 108.91 $\pm$ 2.86 \\
360 & 0.47 $\pm$ 0.26 & 79.41 $\pm$ 2.48 & 8.71 $\pm$ 1.67 & 127.28 $\pm$ 2.88 \\
450 & 0.62 $\pm$ 0.28 & 87.17 $\pm$ 2.25 & 9.21 $\pm$ 1.69 & 127.51 $\pm$ 2.36 \\
540 & 0.66 $\pm$ 0.23 & 96.58 $\pm$ 2.04 & 9.74 $\pm$ 1.53 & 130.59 $\pm$ 1.64 \\
\hline
\end{tabular}
\end{table}

A limitation of our analysis is that the SA implementations and Gurobi were restricted to a standard desktop CPU, while the computational resources leveraged by D-Wave’s hybrid solver are not publicly documented. This asymmetry makes it difficult to establish a fully equitable basis for comparison, since part of the observed performance may arise from classical resources such as high-performance CPUs or GPU acceleration rather than quantum effects. Moreover, the mathematical model implemented in the hybrid solver was formulated exclusively with binary variables, making RS-SA the most direct classical counterpart. While INT-SA systematically outperformed RS-SA, its advantage was constrained by hardware limitations; with access to more powerful classical resources — for example, high-performance CPUs or GPU-accelerated frameworks — INT-SA could potentially achieve even stronger results. Looking ahead, as D-Wave’s hybrid solvers expand to support native integer variables and more general metaheuristic strategies, it may become feasible to incorporate approaches similar to INT-SA into hybrid quantum-classical workflows. 

To complement the benchmark experiments, we conducted a factory-scale simulation spanning approximately three months, using operational data obtained from a real factory. The simulation reproduced the exact warehouse structure, consisting of 496 racks with an average depth of about 12 positions, accommodating around 170 distinct item types. Over this period, we simulated 24,825 insertion and 26,542 removal operations, allowing us to explicitly compute the number of required reinsertion according to each tested strategy. Three allocation methods were compared: (i) the strategy developed in this work (\emph{Proposed Method}); (ii) the factory’s allocation procedure (\emph{Factory’s Method}), providing a qualitative rather than direct comparison as discussed in the following; and (iii) a newly introduced approach termed the \emph{Recommendation Method}, based on the pairwise affinity parameter $\lambda$, enabling a direct comparison to our method since it starts from the same initial stock. The Recommendation Method follows a simple strategy: for each new item, it first checks for racks already dedicated to that item type, and if any exist, it inserts the item; otherwise, it identifies the racks with the smallest $\lambda$ value relative to the item in the last position and inserts the item in the first available one among them. By construction, this approach is limited to sequential insertions ($N=1$).

\begin{figure*}[h]
    \centering
\includegraphics[width=1\linewidth]{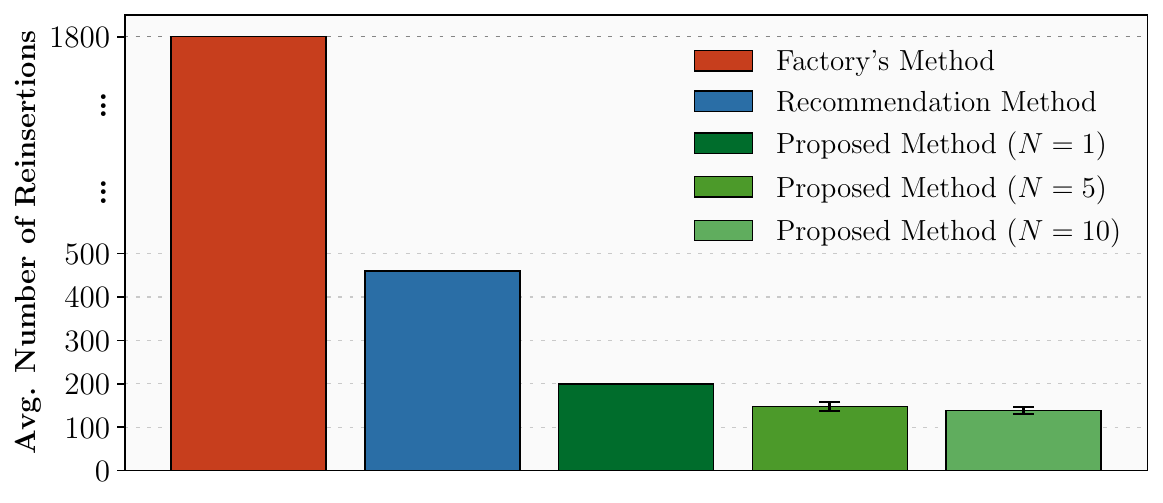}
\caption{
Average number of reinsertions obtained in the factory-scale simulation for the three allocation strategies.
For the Factory’s Method, the reported value corresponds directly to the number of reinsertions recorded in the real factory logs during the three-month period reproduced in the simulation; therefore, no standard deviation is shown.
The Recommendation Method is deterministic by construction.
The Proposed Method yielded the lowest number of reinsertions, with $200 \pm 0$ for $N=1$, $148.2 \pm 9.8$ for $N=5$, and $138 \pm 8.4$ for $N=10$, averaged over eight independent runs.
Error bars indicate standard deviations.}
\label{fig:factory_reinsertions}
\end{figure*}

The average number of required reinsertions obtained in the factory-scale simulation for each allocation method is shown in Fig.~\ref{fig:factory_reinsertions}, with the corresponding means and standard deviations reported in the figure caption. 
For the Factory’s Method, the value displayed corresponds directly to the number of reinsertions observed in the real factory logs during the same period reproduced in the simulation; hence, no standard deviation is presented. The Recommendation Method, in turn, is a deterministic method. The proposed method was evaluated under three scenarios, corresponding to the insertion of 1, 5, and 10 items per decision step, using the integer-variable variant of Simulated Annealing. 
Each configuration was executed eight times independently to estimate the mean and standard deviation. Although the number of items inserted simultaneously may seem modest, the large number of racks results in a combinatorial search space of considerable scale, which justifies the use of metaheuristics.

The results of these large-scale simulations reveal a clear trend: increasing the number of items inserted simultaneously enabled our method to identify globally better placements. When only one item was considered, the resulting decisions were locally optimal but ignored the downstream impact on future insertions. In contrast, jointly evaluating 5 and 10 items, allowed the algorithm to anticipate potential conflicts and distribute items more effectively across shelves, thereby reducing unnecessary reinsertions. This coordinated allocation strategy effectively transformed the problem from a sequence of myopic insertions into an integrated allocation task, resulting in fewer reinsertions and highlighting the advantages of simultaneous allocation.

Compared with the 1801 reinsertions recorded in the factory logs during the same period \parencite{Valerio2025}, the proposed method achieved more than an order of magnitude fewer reinsertions. It is important to note that these factory values include operations starting from an initial inventory state that was not fully documented and therefore had to be reconstructed (see Supplementary Material). For this reason, the reduction cannot be interpreted as a direct one-to-one comparison. Nonetheless, since the simulated period covered multiple full warehouse cycles of filling and emptying, the observed reduction provides a realistic indication of the potential impact of the proposed strategy. Additionally, part of the reinsertions observed in the factory may be attributed to human errors, such as forklift operators not strictly following the allocation instructions. Nevertheless, even when accounting for this factor, the results obtained by the proposed strategy remain robust. In fact, the scenario with $N=1$ already outperforms the factory’s operation, despite being based on sequential allocation. This improvement arises because, in our formulation, the Hamiltonian $H_B$ includes a global term that simultaneously considers all items allocated up to that point and all racks, even when only one item is inserted per step. The parameters of this term are computed from training data, enabling the method to exploit statistical patterns of the system and to anticipate conflicts that would otherwise lead to reinsertion. Thus, the advantage of the approach is not limited to simultaneous allocation with larger $N$, but is already evident in the sequential case.

When compared with the Recommendation Method — which uses the same $\lambda_{\alpha \tau}$ parameters and initial inventory state — the proposed method achieved more than twice the reduction in reinsertions, even for $N=1$. Since both approaches share identical affinity coefficients, this improvement can only be attributed to the formulation of $H_B$, which, in the sequential case, makes the decisions more globally consistent, not by expanding the solution space, but by enriching the information available to each candidate configuration. As $N$ increases, this advantage becomes even more pronounced, as the broader joint view of the items being inserted and their interactions with the items already stored allows the algorithm to anticipate conflicts and distribute them more efficiently across the shelves, thereby explaining the superior results.

These results demonstrate the practical potential of the proposed strategy for real industrial adoption. A complete description of the warehouse model, including the removal procedure, additional operational criteria, and other implementation details, is provided in the Supplementary Material.

\section{Conclusions}\label{SecConclusions}

In this work, we propose a strategy to optimize item allocation in warehouses equipped with gravity-flow racks, where the constraints imposed by the FIFO method can lead to a high number of reinsertions if ineffective strategies are used. Our approach, based on solving a combinatorial optimization problem, prioritizes the joint allocation of items that are frequently requested together. We developed a QUBO Hamiltonian encoding the strategy, enabling implementation on classical, quantum, and quantum-hybrid hardware. 

Through simulations, we compared the performance of D-Wave's CQM Solver with two variants of Simulated Annealing (RS-SA and INT-SA) and with the commercial solver Gurobi across different warehouse configurations. The results clearly show that the D-Wave hybrid solver consistently delivers superior solution quality, with the advantage becoming more pronounced as the problem size increases. Notably, while INT-SA outperforms RS-SA and offers competitive results for small and medium instances, it requires significantly longer runtimes to approach the solution quality achieved by the hybrid solver in larger settings, but still provides a practically viable baseline for medium-scale allocation tasks. Gurobi, in turn, performs well in smaller allocation tasks but shows limited scalability under fixed runtime constraints, with solution quality deteriorating significantly as problem size increases, underscoring its practical scalability limitations.

The findings presented in this work highlight two complementary insights. First, for small- to medium-scale allocation tasks, traditional metaheuristics can provide high-quality solutions within practical runtimes, making them viable options for immediate applications in real factories. Indeed, using real operational data, we adopted INT-SA to demonstrate that our strategy can substantially reduce the number of reinsertions in a production setting. Second, as the number of items and warehouse dimensions scale up, the solution space expands rapidly, and the aforementioned methods face increasing difficulties in escaping local minima. In this regime, D-Wave's CQM Solver provides better-quality solutions in much shorter times, suggesting a clear potential for deployment in large-scale, industrial optimization problems.

As mentioned earlier, D-Wave reports that internal versions of its solvers were tested in two configurations — one with the quantum module activated (\emph{hybrid workflow)} and another with it disabled (\emph{heuristic workflow)} — and that the hybrid configuration consistently converging to better solutions at a faster rate, a behavior referred to as \emph{quantum} or \emph{hybrid acceleration}~\parencite{DWAVE2022}. These internal tests directly support the company’s claims of acceleration and motivate further independent evaluations under real industrial conditions.

The results presented in this work indicate a possible manifestation of quantum acceleration in addressing a practical, real-world optimization problem involving the allocation of hundreds of items within a warehouse. When compared with the best simulated annealing algorithms developed in this work and with a commercial optimization package, the algorithm executed on the CQM Solver achieved superior performance in both solution quality and computation time. However, it is important to emphasize that the underlying mechanisms of D-Wave's hybrid solvers are not available to users, making it unclear whether the quantum module is systematically employed in all runs. Regarding this, D-Wave clarifies that “quantum responses to these queries may be used to guide the heuristic search or to improve the quality of a current pool of solutions”~\parencite{dwavereport}, suggesting that its use may depend on the problem type or internal scheduling protocols. In light of these considerations, although our findings point out to a manifestation of quantum acceleration in a practical, real-scale problem, no definitive statement can be made in this regard, as the D-Wave hybrid solver operates as a closed, proprietary system, preventing direct comparisons between executions with and without the quantum module. In this sense, one of the contributions of this work is to provide an impartial benchmark for assessing the performance of D-Wave’s hybrid solvers, enabling an independent evaluation of classical and quantum-hybrid approaches under comparable conditions \emph{
from a user’s standpoint}, where differences in hardware resources and solver accessibility must be explicitly taken into account.

Beyond comparative benchmarks, the results presented here demonstrate the effectiveness of the proposed strategy, providing a practical method for simultaneously allocating items in gravity-flow racks. Future work should investigate more powerful classical baselines, including GPU-accelerated metaheuristics, integer-variable models for quantum hybrid solvers, and explore multiobjective allocation scenarios, for which the datasets made available in this work can serve as a valuable starting point.

\section*{Acknowledgments}
This work was supported by the Coordenação de Aperfeiçoamento de Pessoal de  Nível Superior (CAPES) - Finance Code 001 and São Paulo Research Foundation (FAPESP) grants No. 2022/00209-6, 2023/04657-6, 2023/14831-3, 2023/18240-0 and 2023/15739-3. C.J.V.-B. is also grateful for the support by the National Council for Scientific and Technological Development (CNPq) Grant No. 311612/2021-0. This work is also part of the MAI/DAI--CNPq Grants No. 139701/2023-0 and 141909/2023-4, CNPq 131088/2022-0 and CNPq 140467/2022-0.

\appendix
\renewcommand{\theequation}{A.\arabic{equation}}
\renewcommand{\thetable}{A\arabic{table}}
\renewcommand{\thefigure}{A\arabic{figure}}

\setcounter{equation}{0}

\setcounter{table}{0}
\setcounter{figure}{0}

\include{supplementary}
\bibliographystyle{elsarticle-harv}
\bibliography{refs4}

\end{document}

%% file: supplementary.tex
\section*{Appendix A – Scaling of the Solution Space Size}
\label{app:SolSp}

This appendix presents a detailed derivation of the solution space size for the proposed allocation problem, taking into account the relevant constraints and assuming that each item to be allocated is distinct. We begin with a simplified case and progressively extend it to the general scenario.

First, consider the case where the number of items to be allocated, $N$, is equal to the total number of available positions, $R = \sum_m R_m$, where $R_m$ denotes the capacity of shelf $m$. The number of ways to assign $R_1$ items to the first shelf is $\binom{N}{R_1}$, leaving $N - R_1$ items to be distributed among the remaining shelves. Similarly, there are $\binom{N-R_1}{R_2}$ ways to allocate the remaining $N-R_1$ items to the second shelf. The total number of feasible allocations up to this point is therefore
$\binom{N}{R_1}\cdot\binom{N-R_1}{R_2}$.
Extending this procedure to all $M$ shelves, the total number of possible configurations is given by
\begin{equation}
    \prod_{m=1}^{M}\binom{N-\sum_{x=1}^{m-1}R_{x}}{R_{m}},
\end{equation} 
\noindent where we define $\sum_{x=1}^{m-1}R_{x}\equiv 0$ for $m=1$.

We now consider the more general case where $N<R$, and the number of items assigned to each shelf, denoted $Q_m$, satisfies $\sum_m Q_m = N$ and $0 \le Q_m \le R_m$. Following the same logic, for a fixed set $\{Q_m\}$, the number of valid allocations is given by
\begin{equation}
\prod_{m=1}^{M}\binom{N-\sum_{x=1}^{m-1}Q_{x}}{Q_{m}}.
\end{equation}

To compute the total number of possible configurations, we must sum this expression over all admissible $M$-tuples $\{Q_m\} \in \mathcal{P}$,
\begin{equation}
\sum_{\{Q_m\}\in\mathcal{P}}\prod_{m=1}^{M}\binom{N-\sum_{x=1}^{m-1}Q_{x}}{Q_{m}},
\label{eq:allPossibilitiesV1}
\end{equation}
where $\mathcal{P}$ denotes the set of all integer vectors ${Q_m}$ satisfying the above constraints.

We can express the summation in Eq.~(\ref{eq:allPossibilitiesV1})
in a way that clarifies how to generate all possible sets of $\{Q_m\}$. Starting with the first shelf, the number of items assigned cannot exceed its capacity $R_1$ or the total number of items $N$. Thus, the maximum number of items that can be placed on it is $\min\{R_1, N\}$. The remaining items can only be allocated if the total space in the other shelves exceeds $N-Q_1 - 1$, \textit{i.e.}, $ \sum_{m=2}^{M}R_m \ge N - Q_1$. This leads to 
\begin{equation}
    Q_1 \ge N - \sum_{m=2}^{M}R_m.
    \label{eq:restrictionQ1}
\end{equation}
If the right-hand side of Eq.~(\ref{eq:restrictionQ1}) is negative, the first shelf may be left empty. Therefore, the lower bound for $Q_1$ is $\max\left\{0, N - \sum_{m=2}^{M}R_m\right\}$.
 
Once $Q_1$ is fixed, $N - Q_1$ items remain to be allocated. The maximum allowable value for $Q_2$ is $\min\left\{{R_2, N - Q_1}\right\}$, while the minimum must ensure that the remaining shelves can fit the rest, leading to a lower bound of $\max\left\{0, N - Q_1 - \sum_{m=3}^{M}R_m\right\}$. By induction, Eq.~(\ref{eq:allPossibilitiesV1}) can be rewritten as
\begin{equation}
        \sum_{Q_1}  \cdots  \sum_{Q_x} \cdots  \sum_{Q_M}     \prod_{m=1}^{M}\binom{N-\sum_{x=1}^{m-1}Q_{x}}{Q_{m}},
        \label{eq:allPossibilitiesV2}
 \end{equation}
\noindent where $Q_x$ ranges from $\max\left\{ 0,N-\sum_{y=1}^{x-1}Q_{y}-\sum_{m=x+1}^{M}R_{m}\right\}$ to $\min\left\{ R_{x},N-\sum_{y=1}^{x-1}Q_{y}\right\}$.

The successive binomial coefficients in Eq.~(\ref{eq:allPossibilitiesV2}) exhibit a telescoping pattern, which allows the expression to be simplified to
\begin{equation}
    \prod_{m=1}^{M}\binom{N-\sum_{x=1}^{m-1}Q_{x}}{Q_{m}}=	\frac{N!}{Q_{1}!}\frac{1}{Q_{2}!}\cdots\frac{1}{Q_{m}!}\frac{1}{\left(N-\sum_{x=1}^{M}Q_{x}\right)!}.
\end{equation}
\noindent Since $\sum_m Q_m = N$, the total number of feasible solutions is given by
\begin{equation}
    S = \sum_{Q_1}  \cdots  \sum_{Q_x} \cdots  \sum_{Q_M}\frac{N!}{\prod_{m=1}^{M}Q_{m}!}.
    \label{eq:restrictionQ2}
\end{equation}

To estimate the scale of the solution space, we compute lower bounds for two warehouse configurations, $10 \times 10$ and $25 \times 25$, where the first number indicates the number of shelves and the second the number of available positions per shelf. For each configuration, we consider the task of allocating $N$ items corresponding to 10--60\% of the total warehouse capacity, consistent with the simulations presented in Sec.~\ref{SecResults}. To simplify the analysis, we calculated as the lower bound the term in the sum of Eq. (\ref{eq:restrictionQ2})
where all shelves contain an equal number of items. When an exact balance was impossible, an additional item was added to selected shelves to achieve uniform distribution.  The resulting lower bounds for each allocation task are shown in Table~\ref{aba}.

\begin{table}[h!]
\centering
\begin{tabular}{|c|c|c|}
\hline
\textbf{Insertion Level (\%)} & \textbf{\(10 \times 10\)} & \textbf{\(25 \times 25\)} \\
\hline
10\% & \(\approx 10^{6}\)   & \(\approx 10^{73}\)  \\
\hline
20\% & \(\approx 10^{15}\)  & \(\approx 10^{157}\) \\
\hline
30\% & \(\approx 10^{24}\)  & \(\approx 10^{243}\) \\
\hline
40\% & \(\approx 10^{34}\)  & \(\approx 10^{328}\) \\
\hline
50\% & \(\approx 10^{43}\)  & \(\approx 10^{415}\) \\
\hline
60\% & \(\approx 10^{53}\)  & \(\approx 10^{501}\) \\
\hline
\end{tabular}
\caption{Lower bounds for the solution space in \(10 \times 10\) and \(25 \times 25\) warehouse configurations under different insertion levels.}
\label{aba}
\end{table}

\section*{Appendix B – Factory-Scale Simulation}

\subsection*{Industrial setting, layout, and data}

To evaluate the performance of the proposed optimization model under realistic industrial conditions, we implemented a large-scale simulation of the functioning of a warehouse spanning a period of approximately three months. The simulation reproduced the actual layout of the warehouse, consisting of 496 gravity-flow racks — 332 with a maximum depth of 13 pallet positions, 7 with 12 positions, and 157 with 10 positions — thereby ensuring that the structural constraints of the facility were faithfully represented. 

The operational cycle modeled in the simulation reflected the daily routines of the factory. Order logs were first read to determine removal requests. The corresponding pallets were then retrieved, and any pallets temporarily displaced to access deeper items were reinserted. The updated warehouse state was subsequently stored as the starting point for the next cycle. Between removal events, insertion operations were performed to replenish stock, closely replicating the process observed in the warehouse during the three-month period. To implement simultaneous allocation, insertion requests were first accumulated in a temporary buffer, which can be interpreted as a staging area within the factory. This approach allowed items to be grouped until the predefined batch size was reached.

Over the three-month period, the simulation processed approximately 24,825 insertion operations and 26,542 removal operations, thereby reflecting the actual workload of the facility. The simulation was initialized with an inventory configuration consistent with the orders observed in the factory logs, ensuring that all requested items could be retrieved and inserted throughout the simulated period. All input data, including rack identifiers, order sequences, and product types, were obtained directly from the factory’s operational records, ensuring that the simulation was grounded in real logistics conditions. Beyond these aspects, data validation confirmed that no orders were missing or invalid. Consequently, all events were replayed strictly in chronological order as specified in the logs.

\subsection*{Dataset and initial inventory reconstruction}

In the original dataset, 50\% of the operations were allocated for training and 50\% for testing, with the division made according to the chronological sequence of insertions and removals. The training set covered the period from 14 March 2024 to 25 April 2024, including all operations registered up to 11:20 a.m. on the latter date. The test set then extended from the subsequent operations on 25 April 2024 through 14 June 2024. Each set comprised a total of 51,367 operations, including both item insertions and removals.

Because no complete log of the warehouse state was available at the beginning of the test period, it was necessary to reconstruct an initial inventory to avoid inconsistencies, particularly in the handling of removal operations. This reconstruction was carried out by identifying items requested for removal that had not appeared in the training-period insertions, under the assumption that such items were already present in the warehouse prior to the start of the dataset. As a consequence, the comparisons with the factory logs should not be interpreted as strictly one-to-one, but rather as indicative of the method’s potential impact.

\subsection*{Training of $\lambda$ parameters}

The training dataset was then submitted to a recommendation algorithm, whose objective was to identify patterns in the removal history and generate association parameters that could be used to minimize future reinsertions. Using the training period already described, we evaluated pairwise removal correlations and encoded them as the matching parameters $\lambda_{\alpha \beta}$, defined in the Hamiltonian formulation. From this complete matrix of $\lambda_{\alpha \beta}$ values and the reconstructed initial stock, we derived the correlations $\lambda_{\alpha \tau}^{(m)}$, which quantify the cost of placing a new item alongside the items already present on that shelf.

\subsection*{Scenarios and simulation protocol}

The simulations were carried out under three scenarios of increasing complexity: a baseline case with $N=1$ (sequential allocation), and two simultaneous allocation cases with $N=5$  and $N=10$. For each scenario, eight independent simulations were performed, where each simulation encompassed the entire three-month period of warehouse operation. 

In every insertion event, the allocation task was solved by executing 100 independent runs of the integer-variable version of Simulated Annealing (INT-SA), implemented in C++. Whenever three solutions of equal energy were obtained across different runs, ties were resolved by selecting the configuration in which items were allocated to racks with greater remaining free space, since layouts with more available capacity imply greater flexibility for subsequent operations and a lower expected number of reinsertions. A stopping criteria were adopted to regulate the optimization process: if no improvement in solution quality was observed over 50 consecutive runs, the best-so-far solution was accepted. 

The removal strategy in the simulation followed the operational practice recommended by the partner factory, prioritizing the retrieval of the item closest to the exit in order to minimize retrieval time. Whenever a requested item was not located in the first position of a rack, the items in front of it were temporarily displaced to a designated buffer shelf (hereafter referred to as a reverse shelf), preserving their order. After the retrieval, these items were reinserted according to the proposed allocation strategy, and the reverse shelf was cleared, ensuring that displaced items were promptly returned to the system, without artificial accumulation outside the warehouse structure.

\subsection*{Computational complexity and algorithmic settings}
Although the tested values of $N$ may appear modest, the combinatorial complexity of the problem is considerable. With 496 racks available, the maximum number of possible allocations scales with $M^N$, where $M$ is the number of racks and $N$ the number of items to be placed. Thus, even for moderate values such as $N=5$, the number of potential configurations grows to the order of $10^{13}$, making exhaustive methods infeasible and justifying the use of metaheuristics such as Simulated Annealing. All simulations employed the integer-variable version of Simulated Annealing (INT-SA), implemented in C++.

The INT-SA algorithm adopted a geometric cooling schedule with cooling factor $\alpha=0.95$, and 
$40\times N$ iterations per temperature level. The initial temperature $T_0$ was defined using a modified version of the method discussed by \textcite{zhan2016list}, in which we average the values obtained from

\begin{equation}
   T= - \frac{\vert \Delta f \vert}{\log p},
   \label{eq:TempFromProb}
\end{equation}
where $\Delta f$ represents the difference between the cost function of the initial state and one of its neighbors, and $p$ is the probability of accepting a transition from a lower to a higher energy. We set $p=0.2$ and averaged over $N$ neighbors. Each run was initialized from a randomized solution, and the stopping criteria were defined either by reaching the maximum number of iterations or by stabilization of the objective function.

\subsection*{Results overview}

The results demonstrate clear benefits from simultaneous allocation. For the baseline case ($N=1$), the simulation produced $200 \pm 0$ reinsertions over the three-month period. With simultaneous allocation of $N=10$, the number of reinsertions further reduced to $138 \pm 8.4$. By contrast, during the same three-month period, the factory logs recorded 1801 reinsertions, i.e., more than an order of magnitude higher than the $N=10$ case. It is important to note, however, that these factory values include operations starting from an initial inventory state that had to be reconstructed (see above). For this reason, the reduction should not be interpreted as a strict one-to-one comparison. Nonetheless, given that the simulated period encompassed multiple complete warehouse cycles of filling and emptying, the observed decrease provides a realistic indication of the practical gains enabled by the proposed strategy.
Moreover, the consistent reduction in reinsertions as 
$N$ increases suggests that simultaneous allocation strategies provide systematic advantages by anticipating potential conflicts and avoiding the limitations inherent in purely sequential decision-making.

\subsection*{Identical items and affinity parameter}

A relevant aspect of the simulation concerns the treatment of identical items. As expected in a real warehouse, the dataset contained a significant number of repeated items. To capture this effect, we extended the affinity parameter $\lambda$, which in our formulation regulates the cost of allocating items to the same rack. For distinct items, affinity values are restricted to the range $[0,1]$, where larger values penalize joint placement due to incompatibility. For identical items, however, the interpretation is inverted: negative values of $\lambda$ provide an energy reduction when items are jointly allocated, thereby encouraging their consolidation. Preliminary tests indicated that $\lambda = -0.5$ achieved the best balance, promoting moderate grouping of identical items without excessive concentration. Sensitivity tests with alternative values of $\lambda$ confirmed that this parameter must be calibrated for each dataset, since its optimal value may vary across factories or operational profiles. A potential refinement of the algorithm would be to determine, already during the training phase in $\lambda_{\alpha \beta}$ and $\lambda_{\alpha \tau}^{m}$ are generated, the optimal value of this parameter for each item, rather than relying on a single global parameter. In the Recommendation Algorithm, the $\lambda$ value for identical items remained zero. Moreover, the algorithm does not sum the parameters of all items stored on a shelf, but rather considers only the one associated with the last inserted position. 

\subsection*{Overall assessment and reproducibility}

The simulation results provide strong evidence for the effectiveness of the proposed allocation strategy. While sequential allocation (\(N=1\)) already outperformed the system currently used in the factory, simultaneous allocation with \(N=5\) and \(N=10\) achieved significantly better results, reducing the number of reinsertions to a fraction of those observed in the factory logs. While these comparisons must be interpreted with caution due to the reconstructed initial inventory, the consistent improvements across scenarios highlight the practical potential of the method to serve as a reliable tool for industrial logistics optimization. Although the source code is not released, all necessary details for replication are provided in this supplementary material, including the warehouse structure, dataset characteristics, algorithmic settings, and evaluation criteria. These specifications ensure that the reported results can be independently reproduced and validated.
